\documentclass[aoas, preprint]{imsart}

\usepackage{amsthm, amsmath, bbm, amssymb}
\usepackage{graphicx}
\RequirePackage[authoryear]{natbib}
\RequirePackage[OT1]{fontenc}
\RequirePackage[colorlinks,citecolor=blue,urlcolor=blue]{hyperref}

\usepackage[textsize=medium]{todonotes}
\usepackage{threeparttable,booktabs}

\startlocaldefs
\newcommand{\order}[1]{{\cal O}\hspace{-0.2em}\left( #1 \right)}
\newcommand\Tstrut{\rule{0pt}{2.6ex}}         
\newcommand\Bstrut{\rule[-0.9ex]{0pt}{0pt}}   
\renewcommand{\Pr}[0]{\mathbb{P}}
\newcommand{\phylogeny}[0]{\mathcal{F}}
\newcommand{\Allparameter}[0]{\boldsymbol{\phi}}
\newcommand{\nTips}[0]{N}

\newcommand{\real}{\rm I\!R}

\newcommand{\Hmat}[1]{\mathbf{H}_{#1}}
\newcommand{\momentum}[0]{\mathbf{p}}
\newcommand{\g}[1]{\mathbf{g}_{#1}}
\newcommand{\s}[1]{\mathbf{s}_{#1}}
\newcommand{\y}[1]{\mathbf{y}_{#1}}
\newcommand{\x}[1]{\mathbf{x}_{#1}}
\newcommand{\I}{\mathbf{I}}
\newcommand{\below}[1]{\mathbf{Y}_{\lfloor #1 \rfloor}}
\newcommand{\abbove}[1]{\mathbf{Y}_{\lceil #1 \rceil}}
\newcommand{\parent}[1]{\mbox{pa}{(#1)}}
\newcommand{\cDensity}[2]{\ensuremath{\Pr(#1 \,|\,#2)}}
\newcommand{\jDensity}[2]{\ensuremath{\Pr(#1 , #2)}}
\newcommand{\p}[1]{\mathbf{p}_{#1}}
\newcommand{\q}[1]{\mathbf{q}_{#1}}
\newcommand{\Ptr}[1]{\mathbf{P}_{#1}}
\newcommand{\gradient}[1]{\frac{\partial}{\partial #1}}
\newcommand{\secondDeriv}[1]{\frac{\partial^2}{{\partial #1}^2}}
\newcommand{\lnP}[1]{\ensuremath{\log \Pr ({#1})}}
\newcommand{\bl}[1]{{b}_{#1}} 
\newcommand{\br}[1]{{r}_{#1}} 
\newcommand{\sr}[1]{{\gamma}_{#1}} 
\newcommand{\nodeT}[1]{t_{#1}} 
\newcommand{\mass}{\mathbf{M}}
\newcommand{\transpose}{^{\prime}}
\endlocaldefs

\begin{document}

\begin{frontmatter}

\title{Gradients \emph{do} grow on trees: a linear-time $\order{\nTips}$-dimensional gradient for statistical phylogenetics}
\runtitle{Phylogenetic gradient-enabled inference}



\begin{aug}
\author{\fnms{Xiang} \snm{Ji}\thanksref{m1}\ead[label=e1]{xji3@ucla.edu}},
\author{\fnms{Zhenyu} \snm{Zhang}\thanksref{m1}\ead[label=e2]{zyz606@ucla.edu}},
\author{\fnms{Andrew} \snm{Holbrook}\thanksref{m1}\ead[label=e3]{aholbroo@ucla.edu}},
\author{\fnms{Akihiko} \snm{Nishimura}\thanksref{m1}\ead[label=e4]{akihiko4@ucla.edu}},
\author{\fnms{Guy} \snm{Baele}\thanksref{m2}\ead[label=e6]{philippe.lemey@kuleuven.be}},
\author{\fnms{Andrew} \snm{Rambaut}\thanksref{m3}\ead[label=e8]{a.rambaut@ed.ac.uk}},
\author{\fnms{Philippe} \snm{Lemey}\thanksref{m2}\ead[label=e7]{guy.baele@kuleuven.be}}
\and
\author{\fnms{Marc A.} \snm{Suchard}\thanksref{m1} \corref{} \ead[label=e9]{msuchard@ucla.edu}}


\runauthor{X.~Ji et al.}

\affiliation{University of California, Los Angeles, USA \thanksmark{m1}}
\affiliation{Department of Microbiology, Immunology and Transplantation, Rega Institute, KU Leuven, Leuven, Belgium \thanksmark{m2}}
\affiliation{University of Edinburgh, Edinburgh, UK \thanksmark{m3}}

\address{
X. Ji\\
A. Holbrook\\
A. Nishimura\\
Department of Biomathematics\\
David Geffen School of Medicine\\
University of California, Los Angeles\\
695 CHARLES E. Young Drive\\
Los Angeles, California 90095-1766\\
USA\\
\printead{e1}\\
\phantom{E-mail:\ }\printead*{e3}\\
\phantom{E-mail:\ }\printead*{e4}}

\address{
Z. Zhang\\
Department of Biostatistics\\
Fielding School of Public Health\\
University of California, Los Angeles\\
Los Angeles, California 90095-1772\\
USA\\
\printead{e2}
}

\address{
G. Baele\\
P. Lemey\\
Department of Microbiology, Immunology and Transplantation\\
Rega Institute\\
KU Leuven\\
Herestraat 49\\
3000 Leuven\\
Belgium\\
\printead{e6}\\
\phantom{E-mail:\ }\printead*{e7}
}

\address{
A. Rambaut\\
Institute of Evolutionary Biology,\\
\, Centre for Immunology, Infection and Evolution\\
University of Edinburgh\\
King's Buildings,\\
Edinburgh, EH9 3FL\\
UK\\
\printead{e8}\\
}

\address{
M. A. Suchard\\
Department of Biostatistics,\\
\, Biostatistics and human genetics\\
University of California, Los Angeles\\
6558 Gonda Building\\
695 CHARLES E. Young Drive\\
Los Angeles, California 90095-1766\\
USA\\
\printead{e9}
}
\end{aug}

\begin{abstract}
Calculation of the log-likelihood stands as the computational bottleneck for many statistical phylogenetic algorithms.
Even worse is its gradient evaluation, often used to target regions of high probability.
Order $\order{\nTips}$-dimensional gradient calculations based on the standard pruning algorithm require $\order{\nTips^2}$ operations where $\nTips$ is the number of sampled molecular sequences.
With the advent of high-throughput sequencing, recent phylogenetic studies have analyzed hundreds to thousands of sequences, with an apparent trend towards even larger data sets as a result of advancing technology.
Such large-scale analyses challenge phylogenetic reconstruction by requiring inference on larger sets of process parameters to model the increasing data heterogeneity.
To make this tractable, we present a linear-time algorithm for $\order{\nTips}$-dimensional gradient evaluation and apply it to general continuous-time Markov processes of sequence substitution on a phylogenetic tree without a need to assume either stationarity or reversibility.
We apply this approach to learn the branch-specific evolutionary rates of three pathogenic viruses: West Nile virus, Dengue virus and Lassa virus.
Our proposed algorithm significantly improves inference efficiency with a $126$- to $234$-fold increase in maximum-likelihood optimization and a $16$- to $33$-fold computational performance increase in a Bayesian framework.
\end{abstract}



\end{frontmatter}

\section{Introduction}

Advances in genome sequencing technology -- that is becoming increasingly portable, accurate and inexpensive (see, e.g., \citealt{quick2016}) -- are generating genetic data at an ever-increasing pace, drastically impacting molecular analyses, from both a statistical and computational perspective.
This is a general challenge in molecular evolution, but the problem is particularly pressing in infectious disease research.
The ability to collect and sequence pathogen genomes in real-time requires the development of novel statistical methods that are able to process the sequences in a timely manner and produce interpretable results to inform national public health organisations, rather than act as a bottleneck to the epidemiological response workflow.
Coupling such methods with highly-efficient computing is key to rapid dissemination of outbreak analysis results to make global health decisions focused on intervention strategies and disease control.
Molecular phylogenetics has become an essential analytical tool for understanding the complex patterns in which rapidly evolving pathogens propagate throughout and between countries, owing to the complex travel and transportation patterns evinced by modern economies \citep{Pybus2015}, along with other factors such as increased global population and urbanisation \citep{bloom2017}.
Of the statistical paradigms employed in this domain, likelihood-based inference is by far the most dominant because of its ability to incorporate complex statistical models while offering accurate tree reconstruction under a wide range of evolutionary scenarios (see, e.g., \citealt{ogden2006}).
These likelihood-based approaches require repeated evaluation of the observed data likelihood function and its gradient and therefore computational performance is heavily dependent on data scale.
As a result, and despite their superior accuracy, faster heuristics often substitute for likelihood-based methods in scenarios where a timely response is essential.

Felsenstein's pruning algorithm \citep{Felsenstein1973, Felsenstein1981} makes the observed data likelihood in phylogenetics computationally tractable.
The observed molecular sequences at the tips evolve on the phylogenetic tree according to a continuous time Markov chain (CTMC) with discrete states.
The pruning algorithm marginalizes over all possible latent states of the CTMC at internal nodes and calculates the probability of the observed sequence data through a post-order tree traversal, which visits all nodes once in a descendant-to-parent fashion that works its way up to the root starting from the tips.
This traversal requires $\order{\nTips}$ operations for each likelihood evaluation, where $\nTips$ is the number of branches.
For a CTMC with discrete states, one can calculate the first derivative of the likelihood by substituting the transition probability matrix with its derivative matrix into the pruning algorithm \citep{Kishino1990, Bryant2005, Kenney2012}.
This pruning-based gradient calculation requires the same computational effort as the likelihood evaluation for a parameter on a given branch, i.e.~$\order{\nTips}$, but costs $\order{\nTips^2}$ operations to calculate with respect to (w.r.t.) parameters pertaining to all branches.
Both maximum-likelihood and Bayesian inference are popular frameworks for inferring the phylogeny and its related evolutionary parameters, requiring the same observed data likelihood to be estimated w.r.t.~the parameter space.
Parameters of interest include the topology of the evolutionary tree, branch lengths, parameters within the infinitesimal generator matrix that describes the CTMC as well as mixture model parameters that describe evolutionary processes such as among-site rate heterogeneity \citep{yang1994maximum} and varying rates between partitions \citep{shapiro2006}.

Owing to the complexity of the phylogenetic likelihood surface (see, e.g., \citealt{sanderson2015}), maximum-likelihood frameworks employ non-linear optimization to find the maximum likelihood estimate (MLE) for model parameters.
Importantly, the computations required to find the MLE differ greatly between parameters, as certain `local' parameters -- often specific to a single branch or a subset of branches -- only require a (small) part of the likelihood function to be re-evaluated whereas other `global' parameters -- typically the parameters of the CTMC process -- require a complete re-evaluation.
In addition to the global optimization routine that re-evaluates the complete likelihood when proposing new parameter values, maximum-likelihood software packages such as RAxML \citep{Stamatakis2004} and GARLI \citep{Zwickl2006} incorporate a local optimization routine that only optimizes a few branch-specific parameters -- e.g.~in the vicinity of a recent topological change -- while keeping all other parameters fixed.
Although both applications adopt pruning-based algorithms for gradient calculations, the computational cost of local optimization routines is roughly only $\order{\nTips}$, which they achieve by optimizing only $\order{1}$ number of parameters, e.g.~the three branch lengths connecting the internal node that is the target of a tree rearrangement operation.
An additional advantage of such local routines is the possibility to perform multiple evaluations of branch-specific derivatives in parallel, conditional on the remainder of the tree not changing.
Bayesian phylogenetic inference packages combine prior knowledge with the (observed data) likelihood into a joint density proportional to the posterior and, as such, attempt to estimate posterior distributions for all parameters of interest.
Despite its great success for incorporating complex statistical models (see, e.g., \citealt{huelsenbeck2001}), Bayesian phylogenetic inference remains computationally intensive.
The computational cost of the gradient evaluation prevents Bayesian phylogenetics from benefiting from more efficient gradient-based samplers, such as the Hamiltonian Monte Carlo (HMC) sampler \citep{neal2011mcmc}.
In summary, both maximum-likelihood and Bayesian implementations of phylogenetic modeling stand to benefit from faster calculations of the gradient.

We here propose an $\order{\nTips}$ algorithm for calculating the gradient w.r.t.~all branch-specific parameters by complementing the post-order traversal in the pruning algorithm with its corresponding pre-order traversal.
The algorithm thus extends the pioneering work of \citet{Schadt1998}
to general CTMCs (homogeneous or not) while not assuming stationarity or reversibility.
We apply our proposed algorithm to study the evolutionary rates of viral sequences that we model with a random-effects clock model that combines both fixed- and random-effects when accommodating evolutionary rate variation \citep{bletsa2019}.
We show that the proposed aproach significantly improves inference efficiency of the branch-specific evolutionary rates under both maximum-likelihood and Bayesian frameworks.

\section{Algorithms} \label{sec:algorithm}
In this section, we define necessary notation for deriving the gradient algorithm.
We then illustrate the likelihood calculation through the post-order traversal as in the pruning algorithm and the update of the post-order partial likelihood vectors.
We derive a new partial likelihood vector at each node and its update through a pre-order traversal.
We expand the likelihood at any node as the inner product of its post- and pre-order partial likelihood vectors.
Finally, we derive the $\order{N}$-dimensional gradient using the two partial likelihood vectors at all nodes.

\subsection{Notation} \label{sec:notations}
Consider a phylogeny $\phylogeny$ with $\nTips$ tips and $\nTips-1$ internal nodes.
Assume that the root node is on the top and the tip nodes are at the bottom of $\phylogeny$.
We denote the tip nodes with numbers $1, 2, ..., \nTips$ and the internal nodes with numbers $\nTips+1, \nTips+2, ..., 2\nTips-1$
where the root node is fixed at $2\nTips-1$.
Any branch on $\phylogeny$ connects a parent node to its child node where the parent node is closer to the root.
We denote $\parent{i}$ as the parent node of node $i$.
We refer to a branch by the number of the child node it connects.
On $\phylogeny$, we model the sites in the sequence alignment as independent and identically distributed such that they arise from conditionally independent CTMCs acting along each branch.
Depending on the state space of the CTMCs, a site can be a single (nucleotide) column or multiple consecutive columns that contain a codon (or encode for an amino acid) or even the entire sequence.

Suppose we have observed (at tips) and latent (at internal nodes) discrete evolutionary characters
$Y_i$ for $i = 1,\ldots, 2\nTips - 1$ at a site.
Character $Y_i$ has $m$ possible states (e.g.~$m=4$ for nucleotide substitution models, $m=20$ for amino acid substitution models and $m=61$ for codon substitution models that exclude the stop-codons).
Let $\bl{i}$ denote the branch length of branch $i$.
Let $\br{i}$ denote the evolutionary rate on branch $i$ and $\nodeT{i}$ denote the real time of node $i$.
Then $\bl{i} = \br{i} (\nodeT{i} - \nodeT{\parent{i}})$.
For branch $i$ with CTMC infinitesimal rate matrix $\mathbf{Q}_i$,
the transition probability matrix is $\Ptr{i} = e^{\mathbf{Q}_i \bl{i}}$.
Let $\boldsymbol{\pi} = \left[ \Pr(Y_{2\nTips-1} = 1), \Pr(Y_{2\nTips-1} = 2), \ldots, \Pr(Y_{2\nTips-1} = m) \right]\transpose$ denote the state distribution at the root node (not necessarily the stationary distribution of the CTMCs).

The evolutionary rates and chronological times appear implicitly in the likelihood function through the branch lengths.
This poses an inference challenge for molecular dating, also known as divergence time estimation.
Having samples with different sampling times, such as serially sampled viral sequences or fossil information, supplements additional time anchors for calibration.
Improvement on characterizing the other confounding factor, the evolutionary rates, relies on the development of more biologically plausible clock models that describe the rate changes on the tree.
However, such models come at the cost of having to infer many highly correlated parameters
that can be computationally demanding
for large datasets (see Section~\ref{clock} for more detail).

To setup the post- and pre-order partial likelihood vectors, we further divide the observed characters
$\mathbf{Y} = \{Y_i, 1 \le i \le \nTips\}$
into two disjoint sets w.r.t.~any node in $\phylogeny$.
Let $\below{i}$ denote the observed characters at the tip nodes that are descendant of node $i$.
Let $\abbove{i} = \mathbf{Y} \setminus \below{i}$ denote the complement set of observed characters.
Finally, let $\Allparameter = \{\phylogeny, \br{i}, \bl{i}, \nodeT{i}, \mathbf{Q}_i; \forall\ i\}$ collect all model parameters.
The length $m$ post-order partial likelihood vector $\p{i}$ of node $i$ at a site has the $j$-th element
being $(\p{i})_j = \cDensity{\below{i}}{Y_i = j}$.
When $i$ is a tip node, $\cDensity{\below{i}}{\mathbf{Y}_i=j} = \mathbbm{1}_{\{\mathbf{Y}_i=j\}}$ for $j=1, 2, \ldots, m$.
For partially observed and missing data at the tip node,
one can modify the post-order partial likelihood vector to reflect this information \citep{Felsenstein1981}.  
Similarly, the pre-order partial likelihood vector $\q{i}$ of node $i$ has the $j$-th element
being $(\q{i})_j = \jDensity{Y_i = j}{\abbove{i}}$.
For the root node,
$\abbove{2\nTips-1} = \emptyset$,
and the pre-order partial likelihood vector is the same as the state distribution (i.e.~$\q{2\nTips-1} = \boldsymbol{\pi}$).

\subsection{Likelihood}

\begin{figure}
\begin{center}
  \includegraphics[width=0.4\textwidth]{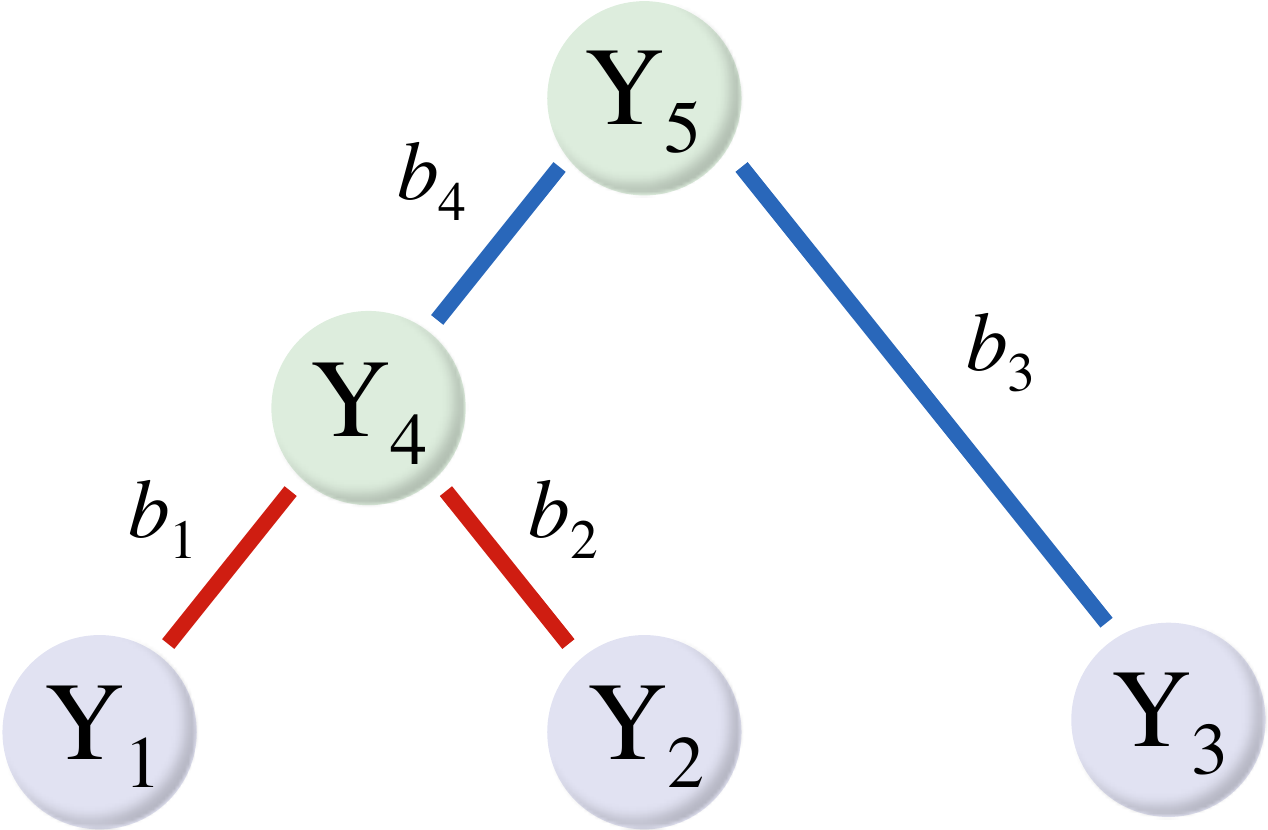}
  \caption{Schematic of a $3$-taxon tree.
  The observed data $\mathbf{Y} = ({\mathbf{Y}_1}, {\mathbf{Y}_2}, {\mathbf{Y}_3})\transpose$ are sequence states at the tips of the tree.
  The latent states ${\mathbf{Y}_4}$ and ${\mathbf{Y}_5}$ are at internal nodes of the tree.
  We divide the observed data $\mathbf{Y}$ into two disjoint sets with
$\below{4} = \{\mathbf{Y}_1, \mathbf{Y}_2\}$ and
$\abbove{4} = \{\mathbf{Y}_3\}$ to help setup the corresponding post- and pre-order partial likelihood vectors at internal node $4$.
  We further color the branches to show the update of the two partial likelihood vectors at internal node $4$ such that
  red branches correspond to the update of the post-order partial likelihood vector
  and
  blue branches correspond to the update of the pre-order partial likelihood vector.}
  \label{fig:exampleTree}
\end{center}
\end{figure}

The likelihood is the marginal probability of the observed discrete characters at the tip nodes that sums over all possible latent characters at the internal nodes:
\begin{equation}
\label{Eq:likelihood}
\begin{aligned}
\Pr{(\mathbf{Y})} &=
\sum\limits_{{\mathbf{Y}_{N+1}}}\sum\limits_{{\mathbf{Y}_{N+2}}} \ldots \sum\limits_{{\mathbf{Y}_{2N-1}}} {\Pr{(\mathbf{Y}, \mathbf{y})}} \text{ and}\\
{\Pr{(\mathbf{Y}, \mathbf{y})}}
&=
  \Pr{(\mathbf{Y}_{2N-1})} \prod\limits_{j=1}^{2N-2} {\cDensity{\mathbf{Y}_j}{\mathbf{Y}_{\parent{j}}}},\\
\end{aligned}
\end{equation}
where the summation at internal nodes are w.r.t.~all possible latent states.
We omit the conditioning on the parameters $\Allparameter$ above and in later derivations to save space.
We use the example phylogenetic tree in Figure~\ref{fig:exampleTree} with $3$ tip nodes and $2$ internal nodes to demonstrate the likelihood calculation.
The observed data in Figure~\ref{fig:exampleTree} are $\mathbf{Y} = \{{\mathbf{Y}_1}, {\mathbf{Y}_2}, {\mathbf{Y}_3}\}$.
And one obtains the likelihood of the observed data by marginalizing over $\mathbf{y}=\{{\mathbf{Y}_4}, {\mathbf{Y}_5}\}$.

\subsection{Post-order traversal}
The pruning algorithm is a dynamic programming algorithm that calculates Equation~\ref{Eq:likelihood} through post-order traversal \citep{Felsenstein1973, Felsenstein1981}.
The post-order traversal visits every node on the tree in a descendent node first fashion.
For example,
two possible post-order traversals for the example tree in Figure~\ref{fig:exampleTree} are
$1 \to 2 \to 3 \to 4 \to 5$ or $1 \to 2 \to 4 \to 3 \to 5$.
%
Using the latter, the decomposition
\begin{equation}
\label{Eq:pruning}
\begin{aligned}
\Pr(\mathbf{Y}) =
&\sum\limits_{{\mathbf{Y}_5}}
\Pr({\mathbf{Y}_5}) \left[ \sum\limits_{{\mathbf{Y}_4}} \cDensity{{\mathbf{Y}_4}}{{\mathbf{Y}_5}} \cDensity{{\mathbf{Y}_1}}{{\mathbf{Y}_4}} \cDensity{{\mathbf{Y}_2}}{{\mathbf{Y}_4}}
   \right] \cDensity{{\mathbf{Y}_3}}{{\mathbf{Y}_5}}
\end{aligned}
\end{equation}
%
%
shows how the pruning algorithm separates the grand sum in Equation~\ref{Eq:likelihood} into intermediate steps at the internal nodes for the example phylogenetic tree.
With the post-order partial likelihood vector and the transition probability matrices,
the matrix-vector representation of Equation~\ref{Eq:pruning} is:
\begin{equation}
\label{Eq:pruningVec}
\begin{aligned}
\Pr(\mathbf{Y}) =
&\boldsymbol{\pi}\transpose  \left[ \Ptr{4} \left( \Ptr{1} \p{1} \circ \Ptr{2} \p{2} \right) \circ \Ptr{3} \p{3}\right],
\\
\end{aligned}
\end{equation}
where $\circ$ denotes the element-wise multiplication. 

Only post-order partial likelihood vectors at the tip nodes appear explicitly in Equation~\ref{Eq:pruningVec}.
The recursive update for the post-order partial likelihood vector $\p{k}$ at internal node $k$ given
the post-order partial likelihood vectors $\p{i}$ and $\p{j}$ at its two descendent nodes $i$ and $j$ (i.e.~$\parent{i} = \parent{j} = k$)
is implicit in Equation~\ref{Eq:pruningVec}:
\begin{equation}
\label{Eq:postOrderPartialUpdate}
\p{k}=\Ptr{i} \p{i} \circ \Ptr{j} \p{j}.
\end{equation}
Again, for the update of the post-order partial likelihood vector at internal node $4$ in Figure~\ref{fig:exampleTree}, this makes $k=4$, $i=1$, $j=2$ and
$\p{4} = \cDensity{\below{4}}{\mathbf{Y}_4} = \Ptr{1} \p{1} \circ \Ptr{2} \p{2}$.
We color the branches relevant to this update red.

The post-order traversal updates all post-order partial likelihood vectors up to the root node.
At the end of the traversal, the likelihood is just the inner product of the state distribution vector with the post-order partial likelihood vector at the root node.
\begin{equation}
\label{Eq:rootInnerProduct}
\Pr(\mathbf{Y})
 = \sum\limits_{j=1}^{m} \left[ \Pr(Y_{2\nTips-1}=j) \cDensity{\below{2\nTips-1}}{Y_{2\nTips-1} = j} \right]
 = \boldsymbol{\pi}\transpose \p{2\nTips-1}.
\end{equation}

In the next section, we expand the likelihood as the inner product at any node of its post- and pre-order partial likelihood vectors.
In fact, this is obvious for the root node because the pre-order partial likelihood vector at the root node is just the state distribution vector and Equation~\ref{Eq:rootInnerProduct} becomes $\Pr(\mathbf{Y})=\q{2\nTips - 1}\transpose \p{2\nTips - 1}$.
The expansion enables us to derive the linear-time algorithm that calculates all branch-specific derivatives at once.

\subsection{Pre-order traversal}


The pre-order traversal starts from the root node, where $\q{2\nTips-1} = \boldsymbol{\pi}$, and updates all remaining pre-order partial likelihood vectors by visiting them in the reverse order of the post-order traversal.
Assume that we have calculated all post-order partial likelihood vectors and consider
recursively internal node $k$ with its two immediate descendent nodes $i$ and $j$.
The pre-order partial likelihood vector for descendent node $i$ falls out as:
\begin{equation}
\label{Eq:preOrderUpdate}
\begin{split}
\jDensity{Y_i}{\abbove{i}}
=& \sum\limits_{Y_k}\jDensity{Y_i, Y_k}{\abbove{k}, \below{j}}\\
=& \sum\limits_{Y_k}\cDensity{Y_i}{Y_k}
{\cDensity{\below{j}}{Y_k} \jDensity{Y_k}{\abbove{k}}}\\
=& \sum\limits_{Y_k}\cDensity{Y_i}{Y_k}
{ \left[ \sum\limits_{Y_j} \cDensity{\below{j}}{Y_j} \cDensity{Y_j}{Y_k} \right]
} \jDensity{Y_k}{\abbove{k}},\\
\end{split}
\end{equation}
since $\cDensity{\below{j}}{Y_j}$ and $\jDensity{Y_k}{\abbove{k}}$ are already known.
The matrix-vector representation of Equation \ref{Eq:preOrderUpdate} is:
\begin{equation}
\q{i} = \Ptr{i}\transpose \left[ \q{k} \circ \left( \Ptr{j} \p{j} \right) \right].
\end{equation}
The derivation of the pre-order partial likelihood vector for node $j$ is similar.
Use Figure~\ref{fig:exampleTree} as an example and consider the update of the pre-order partial likelihood vector at internal node $4$.
Then $i=4$, $j=3$, $k=5$ and
$\q{4} = \Ptr{4}\transpose \left[ \q{5} \circ \left( \Ptr{3} \p{3} \right) \right]$.
We color the branches relevant in this update blue.

For gradient calculations, it becomes useful to rewrite the likelihood as the inner product at any node of
its post- and pre-order partial likelihood vectors.
For node $k$, we have:
\begin{equation}
\label{Eq:forwardBackwardLikelihood}
\begin{aligned}
\Pr(\mathbf{Y})
 &= \sum\limits_{Y_k} \Pr({Y_k, \abbove{k}, \below{k}})\\
 &= \sum\limits_{Y_k} \cDensity{\below{k}}{Y_k} \jDensity{Y_k}{\abbove{k}}\\
 &= \p{k}\transpose \q{k}.\\
\end{aligned}
\end{equation}

In the next section, we derive the derivative of the log-likelihood w.r.t.~any one branch-specific parameter based on Equation~\ref{Eq:forwardBackwardLikelihood}.
In this manner, the new algorithm calculates the gradient of the log-likelihood w.r.t.~all branch-specific parameters at once using $\order{\nTips}$ operations.

\subsection{Gradient}

To ease presentation, we use only the matrix-vector forms for derivation in this section.
The scalar forms are similar to those of the previous sections.
With the likelihood expanded at node $i$ as in Equation~\ref{Eq:forwardBackwardLikelihood}, we derive the gradient vector
of the log-likelihood w.r.t. the branch lengths that has the $i$th element being the partial derivative of the log-likelihood w.r.t. $\bl{i}$:
%
\begin{equation}
\label{Eq:singleRateCaseGradient}
\begin{aligned}
\gradient{\bl{i}}{\lnP{\mathbf{Y}}}
 &=\gradient{\bl{i}}{\left[\p{i}\transpose \q{i} \right]} \bigg/ \Pr(\mathbf{Y})\\
 &=\p{i}\transpose \frac{\partial \q{i}}{\partial \bl{i}} \bigg/ \Pr(\mathbf{Y})\\
 &= \q{i}\transpose \mathbf{Q}_i \p{i} \bigg/ \Pr(\mathbf{Y}),\\
\end{aligned}
\end{equation}
where the third equality follows the fact that the partial derivative
of the pre-order partial likelihood vector $\q{i}$ w.r.t.~the branch length $\bl{i}$ is
%
\begin{equation}\begin{aligned}
\label{Eq:preOrderPartialGradient}
\frac{\partial \q{i}}{\partial \bl{i}}
&= \gradient{\bl{i}}{\left\{ \Ptr{i}\transpose \left[ \q{k} \circ \left( \Ptr{j} \p{j} \right) \right] \right\}} \\
&=  \left( \gradient{\bl{i}}{e^{\mathbf{Q}_i \bl{i}}} \right)\transpose \left[ \q{k} \circ \left( \Ptr{j} \p{j} \right) \right] \\
&=  \left( {e^{\mathbf{Q}_i \bl{i}} \mathbf{Q}_i} \right)\transpose \left[ \q{k} \circ \left( \Ptr{j} \p{j} \right) \right] \\
&= \mathbf{Q}_i\transpose \q{i}.\\
\end{aligned}\end{equation}

\subsection{Likelihood and gradient with substitution rate heterogeneity}

Equation~\ref{Eq:singleRateCaseGradient} assumes homogeneous substitution rate across sites.
A popular approach to model the substitution rate heterogeneity across sites is by using a hidden Markov model
where one models the substitution rate as the discrete hidden state with multiple rate categories \citep{yang1994maximum}.
For discrete rate category $l$ with rate $\sr{l}$, the transition probability matrix for branch $k$ of rate category $l$ is $\Ptr{k|\sr{l}} = e^{\mathbf{Q}_k \bl{k} \sr{l}}$.
As in hidden Markov models, the likelihood becomes the weighted sum of the conditional likelihood of each rate category that marginalizes over all possible hidden states:
%
\begin{equation}
\label{Eq:discreteRateLikelihood}
\begin{aligned}
\Pr(\mathbf{Y})
 &= \sum_{\sr{l}} \cDensity{\mathbf{Y}}{\sr{l}} \Pr(\sr{l})\\
 &= \sum_{\sr{l}} \p{k|\sr{l}}\transpose \q{k|\sr{l}}  \Pr(\sr{l}),\\
\end{aligned}
\end{equation}
where $\p{k|\sr{l}}$ and $\q{k|\sr{l}}$ are the corresponding post- and pre-order partial likelihood vectors at
node $k$ for rate category $l$.
Their updates are the same as in the rate homogeneous case by substituting $\Ptr{k|\sr{l}}$ for $\Ptr{k}$.
Similarly, the numerator and denominator of Equation~\ref{Eq:singleRateCaseGradient} become weighted sums in the rate heterogeneous case:
%
\begin{equation}\label{Eq:GradientWithGamma}
\begin{aligned}
\gradient{\bl{i}}{\lnP{\mathbf{Y}}}
 &=
 {\sum\limits_{\sr{l}} \sr{l} \p{i|\sr{l}}\transpose \mathbf{Q}_i\transpose \q{i|\sr{l}} \Pr(\sr{l})}
  \bigg/ {\Pr(\mathbf{Y})}. \\
\end{aligned}
\end{equation}

Equation~\ref{Eq:preOrderPartialGradient} and Equation~\ref{Eq:GradientWithGamma} show that we only need the post- and pre- order partial likelihood vectors $\p{i}$, $\q{i}$ and the infinitesimal rate matrix $\mathbf{Q}_i$ at node $i$ for calculating the partial derivative of branch $i$.
In fact, we can calculate these matrix-vector multiplications and vector-vector inner products together with the update of the pre-order partial likelihood vectors
in the pre-order traversal.
This gives us the gradient vector of all partial derivatives w.r.t.~branch $1$, $2$, \ldots, $2\nTips - 2$ in one single pre-order traversal.


\subsection{Diagonal elements of the Hessian matrix}

We derive the diagonal elements of the Hessian matrix w.r.t.~the log-likelihood to use it later for preconditioning in section~\ref{sec: HMC}.
The second order derivative of the pre-order partial likelihood vector is similar to that of its gradient by substituting $\mathbf{Q}$ with $\mathbf{Q}^2$ in Equation~\ref{Eq:preOrderPartialGradient}.
Without loss of generality, we illustrate the derivation with the likelihood function in Equation~\ref{Eq:discreteRateLikelihood} where rate homogeneity is its special case with one rate category:
%
\begin{equation}\label{Eq:Hessian}
\begin{aligned}
\secondDeriv{\bl{i}}{\lnP{\mathbf{Y}}}
 &= {\sum\limits_{\sr{l}} \sr{l}^2 {\p{i|\sr{l}}\transpose ({\mathbf{Q}}^2_i)\transpose \q{i|\sr{l}}} \Pr(\sr{l})}
 \bigg/ {\Pr(\mathbf{Y})}  - \left[ \gradient{\bl{i}}{\lnP{\mathbf{Y}}} \right]^2.\\
\end{aligned}
\end{equation}


\subsection{Implementation}

We have implemented a central processing unit (CPU) version of the algorithm in this section in the software package BEAGLE \citep{ayres2019beagle}.
We employ these extensions within the development branch of BEAST \citep{suchard2018} for the demonstrations in this paper.

\section{Applications}

We show that our gradient-based approach significantly improves computational efficiency when drawing inference
with applications in non-linear optimization under a maximum-likelihood framework and through HMC sampling under a Bayesian framework.

\subsection{Non-linear optimization} \label{sec:optimization}

Non-linear optimization is essential to obtain MLEs in statistical phylogenetics.
The parameters include, but are not limited to, branch lengths and substitution rates.
GARLI \citep{Zwickl2006} and RAxML \citep{Stamatakis2004} employ a number of optimization algorithms such as the Newton-Raphson method and Brent's method for various situations.
RAxML can also optionally use the quasi-Newton method of Broyden, Fletcher, Goldfarb, and Shanno, known as the BFGS algorithm (see, e.g., \citealt{dennis1996numerical}), to optimize substitution rate parameters.
The unconstrained optimization of an objective function over a set of real parameters is formulated as:
$\mathop {\min }\limits_{\x{}} f(\x{})$, where $\x{} \in \real^n$ is a real vector with length $n \ge 1$.
In maximum-likelihood inference, the objective function $f: \real^n \to \real$ is the negative log-likelihood.

The past few decades have witnessed the development of a collection of optimization algorithms (see \citealt{nocedal2006numerical, Lange2013} for details).
Here, we revisit the BFGS algorithm and its limited-memory variant (L-BFGS).
We then apply the L-BFGS algorithm for obtaining
the MLE.
All positive parameters in the model are $\log$-transformed into unconstrained parameter spaces.

Like other iterative optimization algorithms, the BFGS algorithm starts at an initial position $\x{0}$ in the parameter space and then iteratively generates a sequence of positions
$\{\x{k}\}_{k=0}^ \infty$.
The BFGS algorithm is a line search method that minimizes
the objective function in each iteration along one specified direction $\boldsymbol{\delta}_k$:
$\mathop {\min }\limits_{\alpha_k > 0} f(\x{k} + \alpha_k \boldsymbol{\delta}_k)$
and the iteration continues at $\x{k+1} = \x{k} + \alpha_k \boldsymbol{\delta}_k$
until iterates make no more fruitful progress, reach a solution point within a certain error tolerance or max out in number of iterations.
Let $\s{k} = \alpha_k \boldsymbol{\delta}_k$ be the increment vector in the parameter space of iteration $k$,
$\g{k} = \nabla f(\x{k})$ be the gradient vector of iteration $k$,
and $\y{k} = \g{k+1} - \g{k}$ be the difference between the gradient vector
of iteration $k + 1$ and the gradient vector of the previous iteration $k$.
BFGS determines the line search direction similarly to that of the Newton method
%
%
except that one approximates the inverse of the Hessian matrix $(\nabla^2 f(\x{k}))^{-1}$ by $\Hmat{k}$:
\begin{equation}
\begin{aligned}
\boldsymbol{\delta}_k &= -\Hmat{k} \g{k} \\
\Hmat{k+1} &= (\I - \rho_k \s{k} \y{k}\transpose) \Hmat{k} (\I - \rho_k \y{k} \s{k}\transpose) + \rho_k \s{k} \s{k}\transpose,\\
\end{aligned}
\label{eq:BFGS-Kk}
\end{equation}
where $\rho_k = \frac{1}{\y{k}\transpose \s{k}}$
and Equation~\ref{eq:BFGS-Kk} satisfies the secant condition $\Hmat{k+1} \y{k} = \s{k}$.
BFGS starts with an `initial' approximate of the inverse Hessian matrix (i.e.~$\Hmat{0}=\Hmat{\text{\tiny init}}$) and updates the $\Hmat{}$ matrix at each iteration.
Alternatively, the L-BFGS algorithm `remembers' only the most recent $m$ iterations such that it initializes $\Hmat{k+1-m}=\Hmat{\text{\tiny init}}$ and applies Equation~\ref{eq:BFGS-Kk} $m$ times to get $\Hmat{k+1}$ for the next iteration.
A typical choice of the initial matrix $\Hmat{\text{\tiny init}}$ is
the product of a scalar constant with the identity matrix (see \citealt{nocedal2006numerical,Lange2013} for choices of the scalar).
Therefore, L-BFGS approximates the Hessian matrix with local curvature information.

\newcommand{\position}{\boldsymbol{\theta}}

\subsection{Hamiltonian Monte Carlo sampling} \label{sec: HMC}
The proposed linear-time gradient algorithm also enables efficient inference under a Bayesian framework through HMC sampling.
HMC is a state-of-the-art Markov chain Monte Carlo (MCMC) method that exploits numerical solutions of Hamiltonian dynamics \citep{neal2011mcmc}.
Given a parameter of interest $\position$ with the posterior density $\pi(\position)$, HMC introduces an auxiliary parameter $\momentum$ and samples from the product density
$\pi(\position, \momentum) = \pi(\position)\pi(\momentum)$.
The parameter $\momentum$ typically follows a multivariate normal distribution $\momentum \sim \mathcal{N}(\mathbf{0},\mass)$ whose covariance matrix $\mass$ is referred to as the `mass matrix.' The basic version of HMC sets the mass matrix to the identity matrix, but we discuss a judicious choice in the next section.

Due to the physical laws that motivate HMC, one refers to
$\position$ as the `position' variable
and $\momentum$ as the `momentum' variable.
One then sets the `potential energy' to the negative log posterior density
$U(\position) = -\log(\pi(\position))$
and the `kinetic energy' to $K(\momentum) = \momentum\transpose \mass^{-1} \momentum/2$.
The sum of the potential and kinetic energy forms the Hamiltonian function
$H(\position, \momentum) = U(\position) + K(\momentum)$.
From the current state $(\position_0, \momentum_0)$, HMC generates a Metropolis proposal \citep{metropolis53}  by simulating Hamiltonian dynamics in
the space $(\position, \momentum)$ that evolves according to the differential equation:
\begin{equation}
\label{eq: HMC}
\begin{aligned}
\frac{\textrm{d} \momentum}{\textrm{d} t} &=  - \nabla U(\position) = \nabla \log \pi(\position) \\
\frac{\textrm{d} \position}{\textrm{d} t} &=  \nabla K(\momentum) = \mass^{-1} \momentum .\\
\end{aligned}
\end{equation}
The popular \textit{leapfrog} method \citep{neal2011mcmc} numerically approximates a solution to Equation~\ref{eq: HMC}. 
Each leapfrog step of size $\epsilon$ follows the trajectory
\begin{equation}
\label{eq:leapfrog}
\begin{aligned}
\momentum_{t + \epsilon/2} &= \momentum_{t} + \frac{\epsilon}{2} \nabla \log \pi(\position_t)\\
\position_{t + \epsilon} &= \position_{t} + \epsilon \mass^{-1} \momentum_{t + \epsilon/2}\\
\momentum_{t + \epsilon} &= \momentum_{t + \epsilon/2} + \frac{\epsilon}{2} \nabla \log \pi(\position_{t + \epsilon}) \, .
\end{aligned}
\end{equation}
We need $n$ leapfrog steps, and hence $n+1$ gradient evaluations, to simulate the dynamics from time $t = 0$ to $t = n \epsilon$.
Such an HMC proposal can have small correlation with the current state, yet be accepted with high probability \citep{neal2011mcmc}.
In particular, HMC promises better scalability in the number of parameters \citep{beskos13} and enjoys wide-ranging successes as one of the most reliable MCMC approaches in general settings \citep{bda13, kruschke14, monnahan16}.



\subsubsection{Preconditioning with adaptive mass matrix informed by the diagonal Hessian} \label{sec:preconditioning}
Geometric structure of the posterior distribution significantly affects the computational efficiency of HMC.
For example, when the scales of the posterior distribution vary among individual parameters, failing to account for such structure may reduce the efficiency of HMC \citep{neal2011mcmc, Stan2017}.
We can adapt HMC for such structure by modifying the dynamics in Equation~\ref{eq: HMC} via an appropriately chosen mass matrix $\mass$.
Replacing the standard identity matrix with a non-identity one is equivalent to  \textit{preconditioning} the posterior distribution via parameter transformation \citep{neal2011mcmc,livingstone14,nishimura16gthmc}.

Practitioners often choose a mass matrix that approximates
the inverse of the posterior covariance matrix of $\theta$ \citep{Stan2017} or the negative Hessian of the posterior distribution \citep{Girolami11}.
These two approaches yield similar mass matrices when the posterior distribution is approximately Gaussian.
For more complex distributions, however, the Hessian better accounts for the underlying geometry \citep{Girolami11}
and is further supported by the linear stability analysis of the leapfrog integrator \citep{Hairer2006}.
Despite its theoretical advantages, a major practical issue with a Hessian-based approach is the obligate use of a $\position$-dependent mass matrix $\mass = \mass\left(\position\right)$.
The corresponding dynamics require computationally demanding numerical integrators, each step of which requires several iterations of evaluating and inverting the mass matrix \citep{Girolami11}.

\newcommand{\iterPerUpdate}{k}
To incorporate information from the Hessian without excessive computational burden, we adaptively tune $\mass$ to estimate the expected Hessian averaged over the posterior distribution.
We further restrict $\mass$ to remain diagonal and hence approximate the diagonals of the expected Hessian only.
This restriction is commonly imposed to regularize the estimate, and a diagonal matrix alone can greatly enhance sampling efficiency of HMC in many situations \citep{Stan2017, pymc16}.
Also, we only update the diagonal mass matrix every $\iterPerUpdate = 10$ HMC iterations so that the cost of computing the expected Hessian diagonals remains negligible.
More precisely, from the first $s$ HMC iterations, we compute
\begin{equation}
\begin{aligned}
H_{ii}^{(s)} &=
	\frac{1}{\lfloor s / \iterPerUpdate \rfloor} \sum_{s \, : \, s / \iterPerUpdate \, \in \, \mathbb{Z}^+}
	\left. - \frac{\partial^2  }{\partial^2 \theta_i} \log \pi(\position) \right|_{\position=\position^{(s)}}
	\\ &
	\approx \mathbb{E}_{\pi(\theta)} \left[
		- \frac{\partial^2}{\partial^2 \theta_i} \log \pi(\position)
	\right].
\end{aligned}
\end{equation}
The $(s+1)^{th}$ iteration then updates the mass matrix with appropriate lower and upper thresholds to make sure that it remains positive-definite and numerically stable:
\begin{equation}
M_{ii}^{(s+1)} = \left\{
	\begin{array}{ll}
	m_{\min} &  \text{if } H_{ii} < m_{\min} \\
	m_{\max} & \text{if } H_{ii} > m_{\max} \\
	H_{ii}^{(s)} & \text{otherwise}
	\end{array} \right.
\end{equation}
for $0 < m_{\min} < m_{\max}$.
The above procedure ensures `vanishing adaptation' $H_{ii}^{(s + 1)} - H_{ii}^{(s)} = \order{s^{-1}}$ such that HMC remains ergodic despite the adaptation \citep{Andrieu2008adap_mcmc}.

\section{Inferring evolutionary rate variation}
\label{clock}

Until the development of the first molecular clock model in the 1960s \citep{zuckerkandl1962molecular, zuckerkandl1965evolutionary}, our understanding of evolutionary time-scale derived mostly from fossil records.
This is because evolutionary rate and time are confounded when comparing homologous DNA sequences.
Molecular clock models provide means to anchor the evolutionary time so that chronological events can be estimated.

\subsection{Molecular clock models}

In its simplest and earliest form, the molecular clock model assumes a constant evolutionary rate across the tree \citep{zuckerkandl1962molecular}.
Researchers often refer to this model as the `strict' clock model.
Over the past few decades, researchers have developed a variety of clock models to accommodate the inadequacy of ignoring rate variation among lineages of the strict clock model (see \citealt{kumar2005, ho2014molecular} for extensive reviews).
One way to characterize a molecular clock model is by the number of unique branch-specific evolutionary rates.
The strict clock model assumes rate homogeneity among all branches.
Multi-rate clock models relax the homogeneity assumption by assigning branches to rate categories.
Branches in the same category share the same evolutionary rate.
The number of categories is usually greater than one but smaller than the total number of branches \citep{hasegawa1989estimation, yoder2000estimation, huelsenbeck2000compound, drummond2010bayesian}.
Relaxed molecular clock models contain the highest possible number of unique branch-specific rates where each branch evolves at its own rate.
There are two major classes of relaxed molecular clock models, autocorrelated and uncorrelated clock models.
The major difference between the two classes is their assumption about the causation of the rate variation.
Autocorrelated relaxed clock models assume that evolutionary rate undergoes a diffusion process from the root node to successive branches \citep{thorne1998estimating, kishino2001performance, aris2002effects}, whereas uncorrelated clock models make no assumption of rate correlation among branches \citep{drummond2006relaxed, rannala2007inferring, lemey2010phylogeography}.
A recent addition to the growing list of clock models consists of a mixed relaxed clock model that combines the merits of autocorrelated and uncorrelated relaxed clocks \citep{lartillot2016mixed}.

Application of relaxed clock models inevitably leads to higher dimensional parameter spaces.
However, the computational efficiency of existing methods limits our ability to draw likelihood-based inference from these high-dimensional evolutionary models, a problem that is exacerbated in large data sets.
We show that our new gradient algorithm ameliorates this difficulty through applications in gradient-based optimization methods and HMC sampling.
Specifically, we demonstrate marked improvement on computational efficiency for inferring the evolutionary rates of three viruses under a random-effects relaxed clock model.

\subsection{Random-effects relaxed clock models}

The random-effects relaxed clock model combines a strict clock and an uncorrelated relaxed clock model.
We model the evolutionary rate $\br{i}$ of branch $i$ as the product of a global tree-wise mean parameter $\mu$ and a branch-specific random effect $\epsilon_i$.
We model the random effect $\epsilon_i$'s as independent and identically distributed from a lognormal distribution such that $\epsilon_i$ has mean $1$ and variance $\psi^2$ under a hierarchical model where $\psi$ is the scale parameter.
We note that the popular uncorrelated relaxed clock model is a special case of this clock model and will hence also benefit from the improvements in this paper.

\subsection{Priors}

We assign a conditional reference prior to the global tree-wise mean parameter $\mu$ \citep{Ferreira2008} and an exponential prior with mean $\frac{1}{3}$ to the scale parameter $\psi$.
%
We use the same substitution models 
as in each example's original study \citep{Pybus2012, andersen2015clinical, nunes2014air}.
We provide the BEAST XML files for these analyses in Supplementary Materials.

\subsection{Emerging viral sequences}

We examine the molecular evolution of West Nile virus (WNV) in North America (1999 - 2007), the S segment of Lassa virus (LASV) in West Africa (2008 - 2013) and serotype $3$ of Dengue virus (DENV) in Brazil (1964 - 2010) \citep{Pybus2012, andersen2015clinical, nunes2014air}.
In all three virus data sets, phylogenetic analyses have revealed a high variation of the evolutionary rates across branches in the underlying phylogeny.

\paragraph{West Nile virus}

WNV is a mosquito-borne RNA virus with birds as the primary host.
The first detected case in the United States was in New York City in August 1999, and the virus reached the American west coast by 2004.
In total, human infections resulted in over 1,200 deaths.
The data consist of 104 full genomes, with a total alignment length of 11,029 nucleotides, and were collected from infected human plasma samples from 2003 to 2007 as well as near-complete genomes obtained from GenBank \citep{Pybus2012}.

\paragraph{Lassa virus}

Every year, LASV is responsible for thousands of deaths and tens-of-thousands of hospitalisations \citep{andersen2015clinical}.
While many LASV infections are subclinical, they can also lead to Lassa fever, a hemorrhagic fever similar to that caused by Ebola virus.
Perhaps less well-known than Ebola viral disease, Lassa fever can nonetheless lead to over 50\% fatality rates among hospitalised patients.
Unlike Ebola virus, which passes directly between humans, LASV circulates in a rodent (\emph{Mastomys natalensis}) reservoir and mainly infects humans through contact with rodent excreta.
LASV is a single-stranded RNA virus with a genome consisting of two segments: the L segment is $7.3$ kilobase pairs (kb) long; the S segment is $3.4$ kb long.
In this paper, we use the S segment of the LASV sequence data set of \citet{andersen2015clinical} that consists of $211$ samples obtained at clinics in both Sierra Leone and Nigeria, rodents in the field, laboratory isolates and previously sequenced genomes.

\paragraph{Dengue virus}

Worldwide, DENV infects close to $400$ million people and causes more than $25,000$ deaths annually.
Much like the LASV, DENV can also lead to hemorrhagic fever that is often referred to as `breakbone fever' on account of the severe joint and muscle pain it causes.
DENV is endemic to the tropics and sub-tropics, with mosquitoes transmitting the virus between humans.
\citet{nunes2014air} selected $352$ serotype $3$ DENV (DENV-$3$) sequences from a total of $639$ complete DENV genomes based on genetic diversity and maximization of the sampling interval.
The sample collection ranged from $1964$ to $2010$ within a total of $31$ distinct countries in Southeast Asia, North America, Central America, the Caribbean, and South American countries.


\section{Results}

We present the computational efficiency improvements conferred by our linear-time gradient algorithm for inferring the branch-specific evolutionary rates in this section.

\subsection{Optimization}

We obtain  MLEs of all branch-specific evolutionary rates via the L-BFGS algorithm for all three viral datasets.
In computing these MLEs, we compare the performance of our analytic gradient method with an often-used central finite difference scheme.
The numerical scheme calculates the partial derivative of one branch-specific rate through two likelihood evaluations and has a complexity of $\order{\nTips^2}$ for the gradient w.r.t.~all rates.
On the other hand, our analytic approach scales $\order{\nTips}$ (see Section~\ref{sec:algorithm}).
Table~\ref{tab:MLE_Speedup} shows a summary of the comparison, illustrating the immense performance increase across the three data sets of our analytic method.
Averaged over each iteration of the MLE estimation process, the analytic method outperforms the finite difference scheme by a factor of 126- to 235-fold, leading to a total real time speed up of 210- to 253-fold.

\begin{table}
\caption{
Maximum likelihood estimate (MLE) inference efficiency using two optimization methods: our proposed gradient method (Analytic) and a central finite difference numerical scheme (Numeric).
For each example and method, we report the total time to complete MLE inference, as well as the number of iterations required for optimization on an Intel Core i7-2600 quad-core processor running at 3.40 Ghz.
Our proposed method yields a minimum 200-fold increase in performance across the entire inference, which averages out to a minimum 126-fold performance increase per iteration.
}
\label{tab:MLE_Speedup}
\centering
\resizebox{\textwidth}{!}{\begin{tabular}{lcrcrccc}
\toprule
& & \multicolumn{2}{c}{Analytic} & \multicolumn{2}{c}{Numeric} & \multicolumn{2}{c}{Speedup}\\
\cmidrule(lr){3-4} \cmidrule(lr){5-6} \cmidrule(lr){7-8}
\multicolumn{1}{c}{Example} & \multicolumn{1}{c}{\# Rates} & Time(s) & Iterations  & Time(s) & Iterations & per Iteration & Total \\
\hline
WNV       & 206 & 0.3  & 12 &  59.3  & 20 & 126.2$\times$ & 210.4$\times$ \Tstrut\\
LASV      & 420 & 1.2  & 10 & 369.1  & 19 & 168.8$\times$ & 320.6$\times$ \strut\\
DENV      & 702 & 19.1 & 90 & 4827.9 & 97 & 234.8$\times$ & 253.1$\times$ \Bstrut\\
\bottomrule
\end{tabular}}
\end{table}




\subsection{Posterior inference}

We infer the posterior distribution of all evolutionary rates using three different MCMC transition kernels in BEAST using BEAGLE.
The first transition kernel is the univariate transition kernel that \citet{Pybus2012} formerly employed, which we will refer to as `Univariate'.
`Univariate' updates propose new values for one rate $\br{i}$
at a time whereas the HMC transition kernels propose new values for all $2\nTips - 2$ rates simultaneously.
We consider two mass matrix choices for HMC.  `Vanilla' HMC (vHMC) employs an identity matrix and `preconditioned' HMC (pHMC) employs an adaptive diagonal matrix informed by the Hessian.

We compare the efficiency of these three transition kernels through their effective sample size (ESS) per unit time for estimating all branch-specific evolutionary rates.
For each analysis, we fix the number of MCMC iterations such that they run for approximately the same time, i.e.~100,000 iterations for both HMC kernels compared to 15 million iterations for the univariate kernel when analysing the WNV data set, 50,000 iterations for both HMC kernels compared to 20 million iterations for the univariate kernel when analysing the LASV data set, and 20,000 iterations for both HMC kernels compared to 7.5 million iterations for the univariate kernel when analysing the DENV data set.
\par
\begin{figure}
\begin{center}
  \includegraphics[scale=0.45]{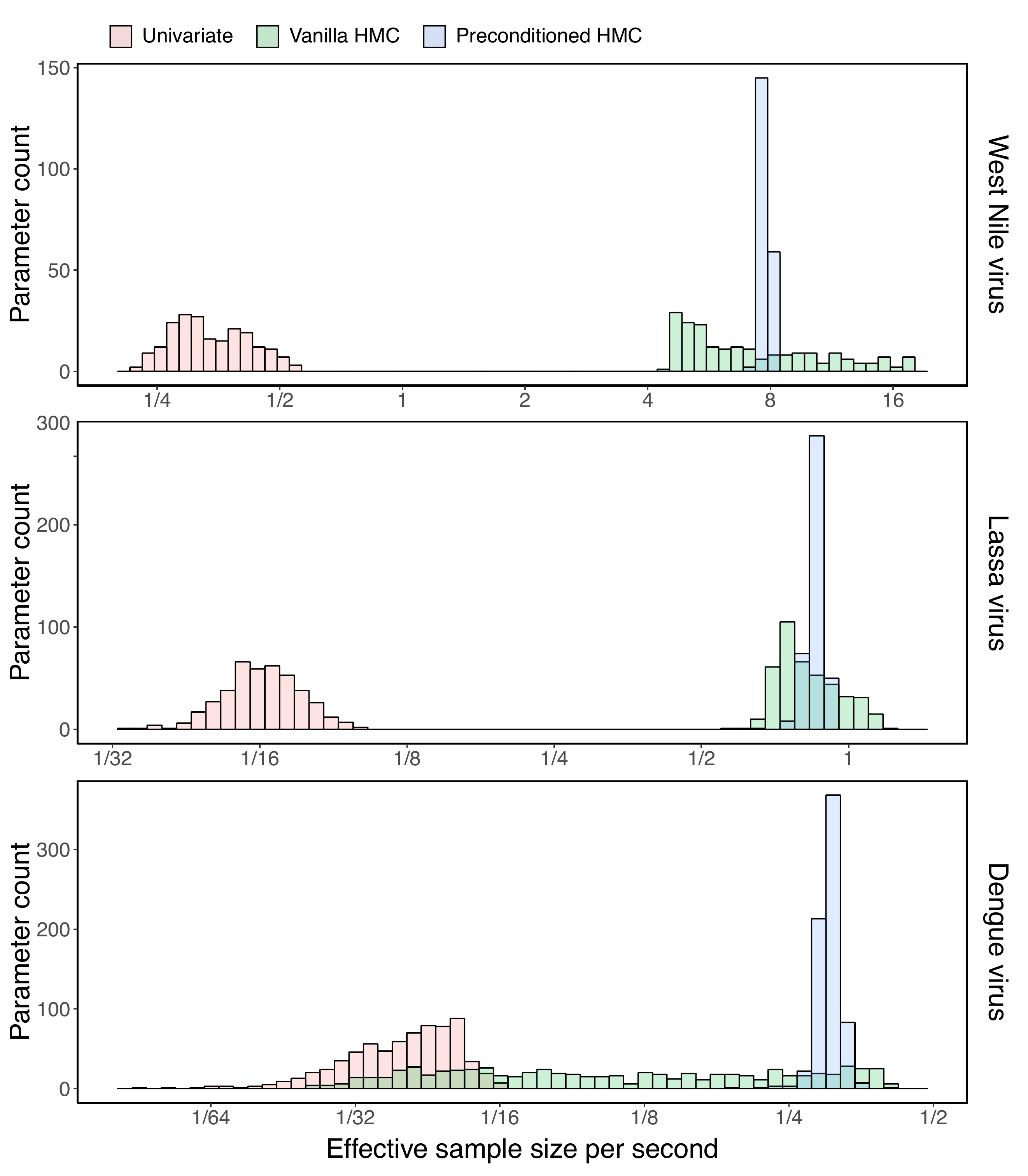}
\end{center}
\caption{
Posterior sampling efficiency on all branch-specific evolutionary rate for the West Nile virus, Lassa virus and Dengue virus examples.
We bin parameters by their ESS/s values.
The three transition kernels employed in the MCMC are color-coded: a univariate transition kernel, a `vanilla' HMC transition kernel with an identity mass matrix and a `preconditioned' HMC transition kernel with an adaptive mass matrix informed by the diagonal elements of the Hessian matrix.
}
\label{fig:ESS}
\end{figure}

Figure~\ref{fig:ESS} illustrates the rate estimates binned by their ESS per second for the three virus data sets, and
table~\ref{tab:ESS} reports the relative increase in ESS per second of the two HMC samplers compared with the univariate kernel over all branch-specific evolutionary rates.
Compared with the univariate kernel, the vHMC sampler achieves a 2.2- to 20.9-fold speedup, whereas the pHMC sampler achieves a 16.4- to 33.9-fold speedup in terms of the minimum ESS per unit time.
The vHMC sampler achieves a 2.5- to 19.8-fold speedup in terms of the median ESS per unit time, whereas the pHMC sampler achieves a 7.4- to 23.9-fold speedup.
The unusual spread of the ESS per second distribution for the vHMC sampler under the Dengue virus example is likely attributable to large variation among the scales of the branch-specific evolutionary rates as discussed in more detail in Section~\ref{sec:discussion}.
The more uniform sampling efficiency of the pHMC sampler arises from the accommodation of the variability in scales among the rates in the mass matrix. 

\begin{table}
\caption{
Relative speedup in terms of effective sample size per second (ESS/s) of our `vanilla' HMC (vHMC) and `preconditioned' HMC (pHMC) transition kernels over a univariate transition kernel, for all three virus data sets.
We report speedup with respect to the minimum and median ESS/s across parameters for each example and method.
}
\label{tab:ESS}
\centering
\begin{tabular}{lcccccc}
\toprule
& \multicolumn{2}{c}{WNV} & \multicolumn{2}{c}{LASV} & \multicolumn{2}{c}{DENV}\\
\cmidrule(lr){2-3} \cmidrule(lr){4-5} \cmidrule(lr){6-7}
\multicolumn{1}{r}{} & vHMC & pHMC  & vHMC & pHMC & vHMC & pHMC \\
\hline
minimum           & 20.9$\times$ & 33.9$\times$ & 16.7$\times$ & 19.8$\times$ & 2.2$\times$ & 16.4$\times$ \Tstrut\\
median        & 19.8$\times$ & 23.9$\times$ & 12.6$\times$ & 13.6$\times$ & 2.5$\times$ & 7.4$\times$ \Bstrut\\
\bottomrule
\end{tabular}
\end{table}

\subsection{Rates of molecular evolution}

We use BEAST in combination with BEAGLE to infer the branch-specific evolutionary rates of the three virus examples under a random-effects relaxed clock model.
The BEAST analyses comprise 20 million MCMC iterations for the WNV data set, 10 million iterations for the LASV data set and 60 million iterations for the DENV data set, to achieve sufficiently high ESS values for all branch-specific evolutionary rates, as assessed using Tracer \citep{rambaut2018}.
In accompanying inferred phylogeny figures,
we color the branches according to their inferred posterior mean branch-specific evolutionary rate.
The range of colors reflects the high variation of rates in all three virus examples.

\paragraph{West Nile virus}

\begin{figure}
\begin{center}
\includegraphics[width=\linewidth]{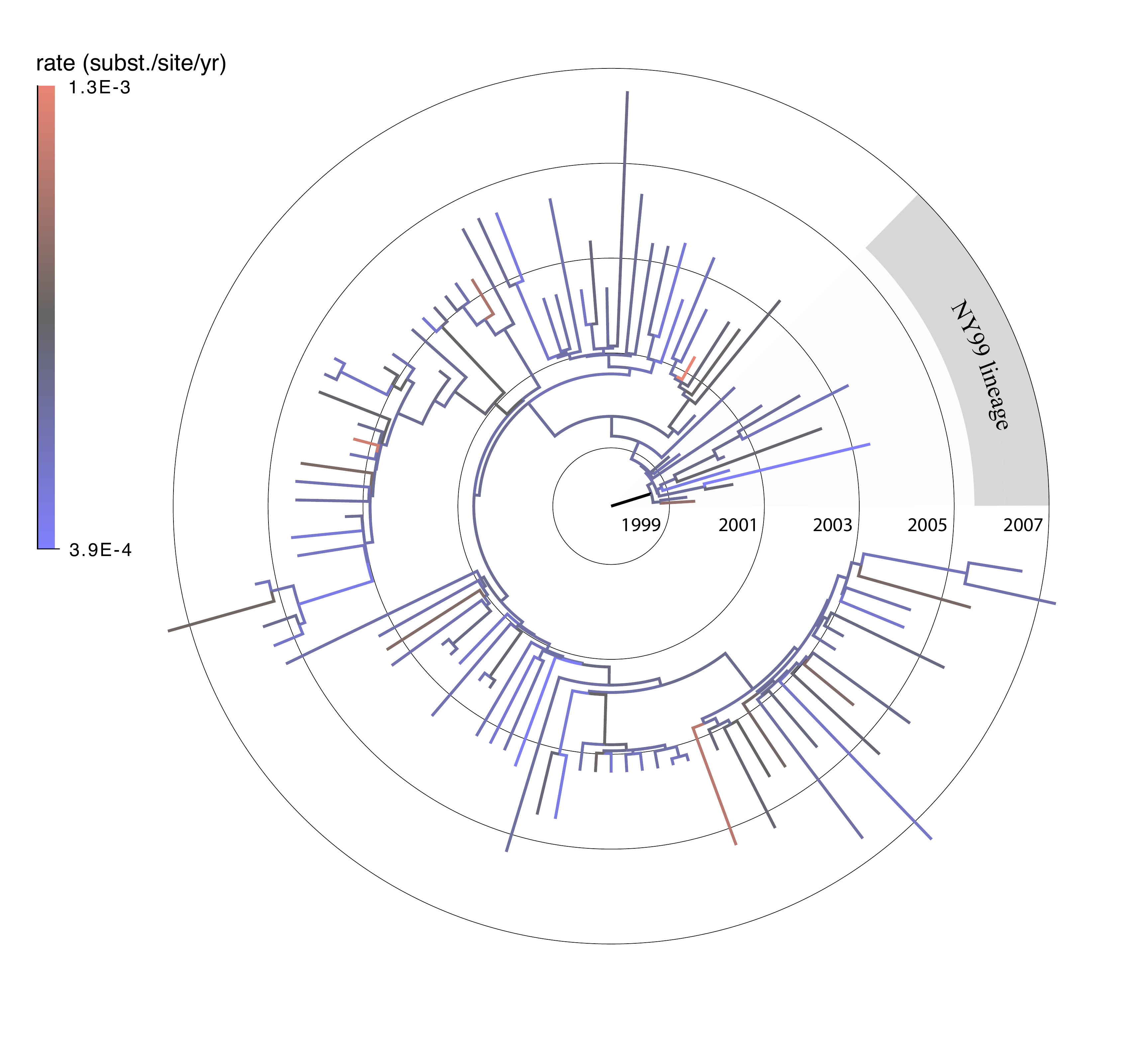}
\end{center}
\caption{
Maximum clade credible tree of the West Nile virus example.
The dataset consists of $104$ sequences of the West Nile virus.
Branches are color-coded by the posterior means of the branch-specific evolutionary rates.
The concentric circles indicate the time scale with the year numbers.
The grey sector in the outer ring indicates the same $13$ samples of the NY99 lineage as identified in the original study.
}
\label{fig:WNV_tree}
\end{figure}

Our analysis estimates the tree-wise (fixed-effect) mean rate $\mu$ with posterior mean $5.67$ (95\% Bayesian credible interval: $5.04, 6.30$) $\times 10^{-4}$ substitutions per site per year
%
%
with an estimated variability characterized by scale $\psi$ with posterior mean $0.33$ $(0.21, 0.46)$ similar to previous estimates \citep{Pybus2012}.
Figure~\ref{fig:WNV_tree} shows the maximum clade credible evolutionary tree of the WNV example.
Our analysis discriminates the NY99 lineage as defined in \cite{davis2005phylogenetic}.
The NY99 lineage is basal to all other genomes congruent with the American epidemic likely to result from the introduction of a single highly pathogenic lineage.

\begin{figure}
\begin{center}
\includegraphics[width=\linewidth]{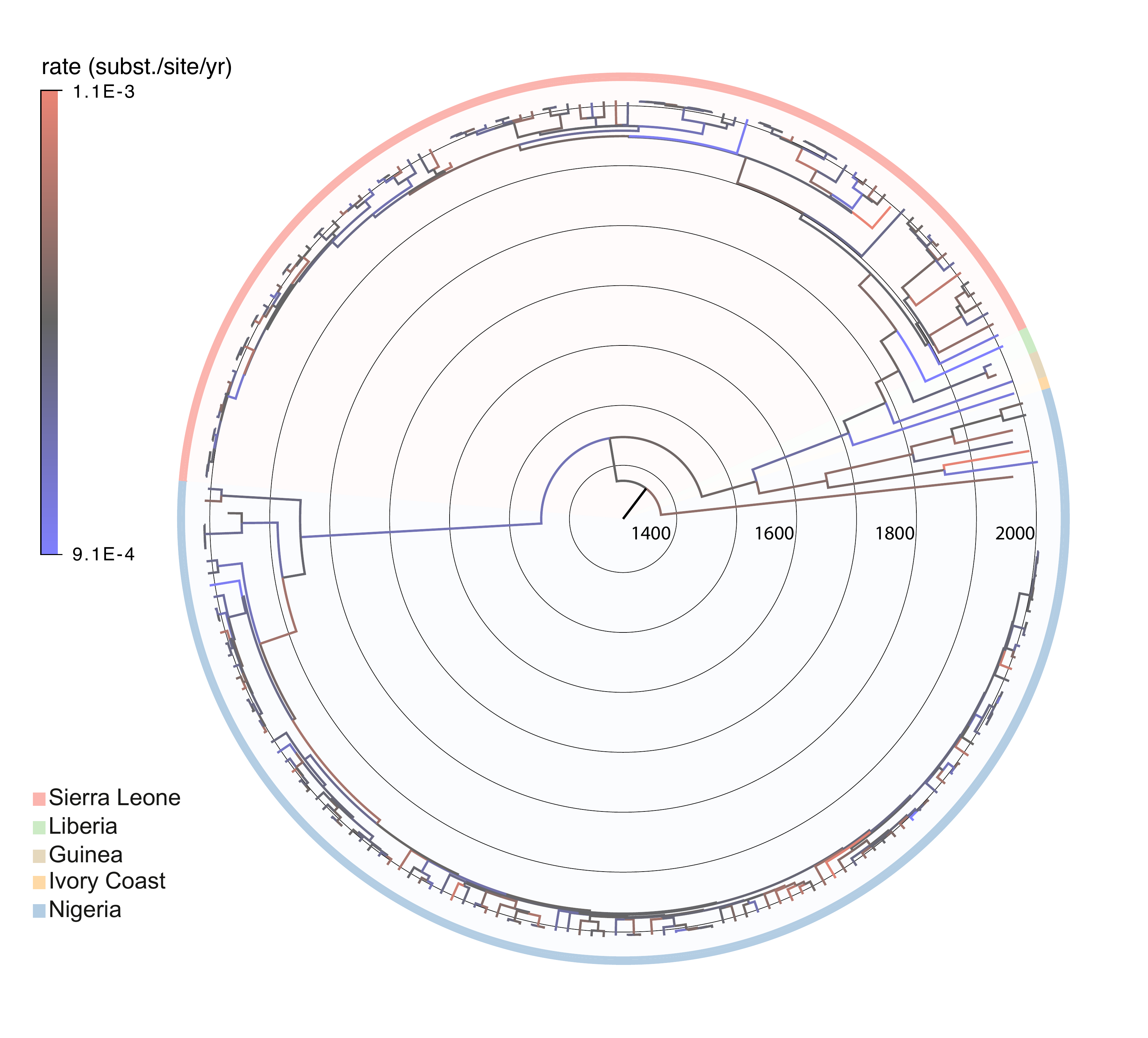}
\end{center}
\caption{
Maximum clade credible tree of the Lassa virus example.
The dataset consists of $211$ sequences of the S segment of the Lassa virus.
Branches are color-coded by the posterior means of the branch-specific evolutionary rates according to the color bar on the top left.
The concentric circles indicate the time scale with the year numbers.
The outer ring indicates the geographic locations of the samples by the color code on the bottom left.
}
\label{fig:Lassa_tree}
\end{figure}

\paragraph{Lassa virus}
Our analysis estimates $\mu = 1.00$ ($0.97, 1.10$) $\times 10^{-3}$ substitutions per site per year
for the S segment of LASV similar to previous estimates \citep{andersen2015clinical, kafetzopoulou2019metagenomic}, with more rate variability ($\psi = 0.088 [0.029, 0.142]$) as compared to WNV.
Figure~\ref{fig:Lassa_tree} shows the maximum clade credible evolutionary tree of the LASV example.
Our result agrees with LASV being a long-standing human pathogen that likely originated in modern-day Nigeria more than a thousand years ago and spread into neighboring West African countries within the last several hundred years \citep{andersen2015clinical, kafetzopoulou2019metagenomic}.

\begin{figure}
\begin{center}
\includegraphics[width=\linewidth]{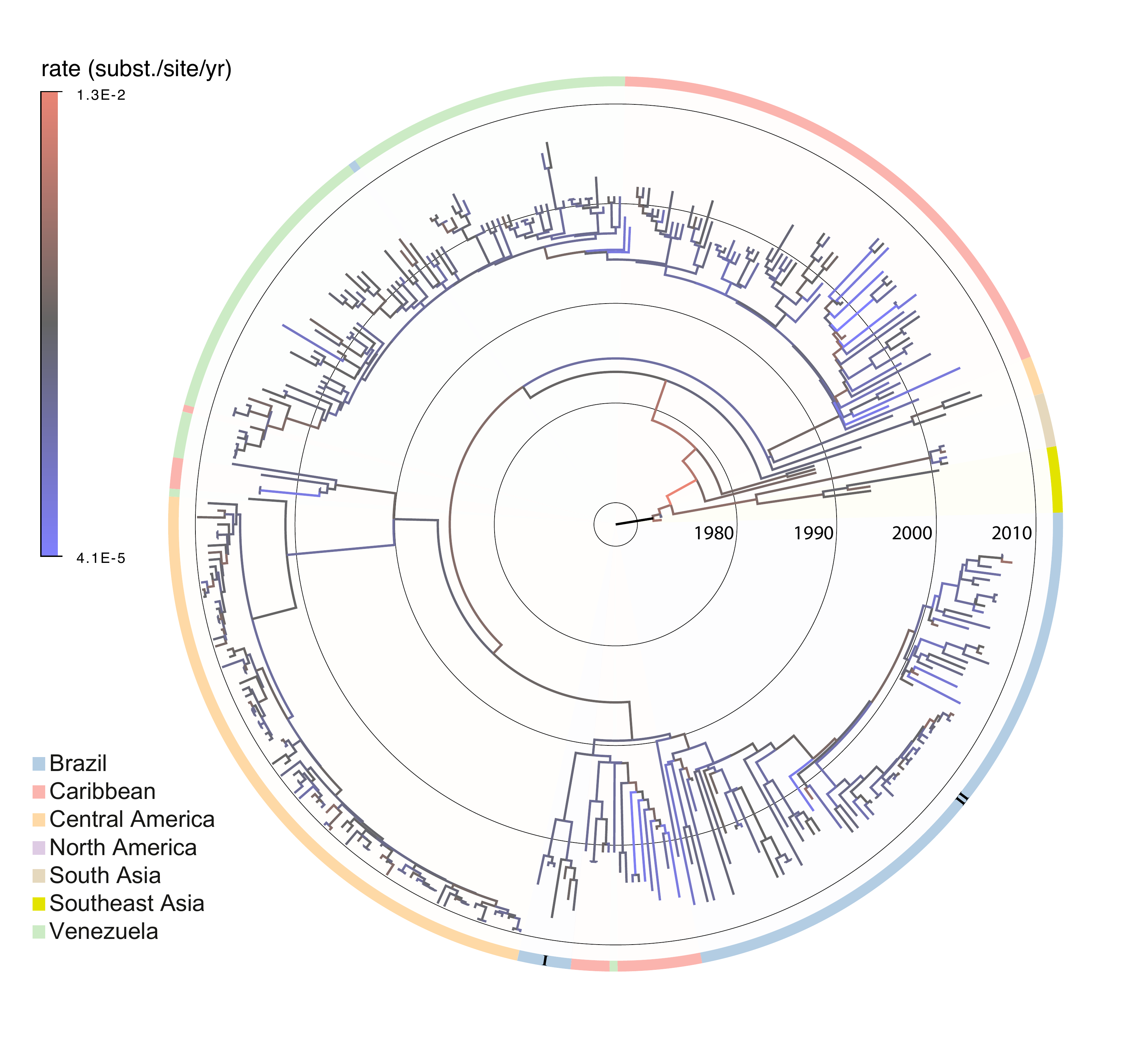}
\end{center}
\caption{
Maximum clade credible tree of the Dengue virus example.
The data set consists of $352$ sequences of the serotype $3$ of the Dengue virus.
Branches are color-coded by the posterior means of the branch-specific evolutionary rates according to the color bar on the top left.
The concentric circles indicate the time scale with the year numbers.
The outer ring indicates the geographic locations of the samples by the color code on the bottom left.
`\textbf{I}' and `\textbf{II}' indicate the two Brazilian lineages as in the original study.
}
\label{fig:Dengue_tree}
\end{figure}

\paragraph{Dengue virus}
Our analysis estimates $\mu = 4.75$ ($4.05, 5.33$) $\times 10^{-4}$ substitutions per site per year
for serotype $3$ of DENV similar to previous estimates \citep{allicock2012phylogeography, nunes2014air}, with the largest rate variability of all examples analysed here ($\psi = 1.26 [1.06, 1.45]$).
Figure~\ref{fig:Dengue_tree} shows the maximum clade credible evolutionary tree of the DENV example.
We identify the same two Brazilian lineages as in \cite{nunes2014air}, and both lineages appear to originate from the Caribbean.



\section{Discussion} \label{sec:discussion}
We presented a new algorithm for evaluating the gradient of the phylogenetic model likelihood w.r.t.~branch-specific parameters.
Our approach achieves linear complexity in the number of sequences by complementing the post-order traversal in Felsenstein's pruning algorithm \citep{Felsenstein1973, Felsenstein1981} with its reverse pre-order traversal.
The two traversals together complete Baum's forward-backward algorithm \citep{Baum1972}.
\cite{Schadt1998} previously employed the forward-backward algorithm to calculate the likelihood and its gradient w.r.t.~the relatively small number of parameters that characterize a generalized \citet{Kimura1980} CTMC.
On the other hand, pruning-only-based gradient algorithms have made improvements over the past few years that scale $\order{\nTips h}$ instead of $\order{\nTips^2}$ where $h$ is the total level of the tree \citep{Kenney2012}.
However, in many phylogenetic problems with non-neutral evolutionary processes, $h$ is often much closer to $\nTips$ than $\log \nTips$.
Careful reuse of some computations when properly re-rooting the tree can further accelerate the pruning-based gradient method.
Unfortunately, re-rooting the tree requires the CTMC to be time-reversible and at stationarity.
The assumptions of reversibility and stationarity can be biologically unreasonable but are often kept for simplicity and computational tractability.
Our linear-time gradient algorithm extends the approach in \cite{Schadt1998} to general CTMCs.
Our algorithm does not require any model assumptions on stationarity or reversibility
and can be applied to both homogeneous and non-homogeneous Markov processes.

Our algorithm calculates the likelihood and its gradient w.r.t.~all branch-specific parameters through the post-order and the complementary pre-order traversal.
One essential benefit of the proposed algorithm is that it calculates the gradient w.r.t.~a collection of branch-specific parameters (e.g.~evolutionary rate and time parameters) at the same time with no additional cost for caching.
However, the computational load is not identical for the two traversals.
For example, the post-order traversal calculates the transition probabilities at all branches that can be reused in the pre-order traversal (see Equation~\ref{Eq:singleRateCaseGradient} and Equation~\ref{Eq:preOrderPartialGradient}).
Moreover, the pre-order traversal updates approximately twice as many partial likelihood vectors as the post-order traversal.
This is due to the additional pre-order partial likelihood vectors at the tip nodes together with the post- and pre-order partial likelihood vectors at the internal nodes.

Through our three example datasets, we illustrate the use of our gradient algorithm in both maximum-likelihood and Bayesian analyses.
We show that our new algorithm can considerably accelerate inference in both frameworks.
In the maximum-likelihood analyses, we compare the performance of the L-BFGS optimization method using our gradient algorithm with
the same optimizer but using a central finite difference numerical gradient algorithm.
We choose this numerical scheme for two reasons.
One is that the central scheme has only roughly twice the computational cost as pruning-based analytical gradient methods.
The other reason is to investigate the influence of numerical error in optimization.
The observed per-iteration speedup with our gradient algorithm increases with increasing number of sequences in the dataset.
This is consistent with our gradient algorithm being a linear-time algorithm in the number of sequences as opposed to quadratic pruning-based algorithms.
We also observe slightly more iterations in the optimization with the numeric gradient than with the proposed analytic gradient method.
Moreover, for all three datasets, the optimization with our analytic gradient method ends with slightly higher log-likelihood values at the $5^{th}$ digit after the decimal point with the same stopping criteria.
The ${\ell}^2$-norm of the gradient when the optimization stops is higher with the numerical method suggesting early termination due to numerical trouble.
Numerical error builds up from the matrix exponential calculations and propagates along the tree.

A caveat of our optimization comparison is that we do not compare with other widely used optimization criteria.
For example, GARLI \citep{Zwickl2006} and RAxML \citep{Stamatakis2004} incorporate local optimization routines
in addition to global optimization.
The purpose of local optimization is partly to avoid the computational burden of optimizing all branches simultaneously, especially after a topological rearrangement.
For time-reversible models at stationarity, with properly rerooting the tree, the branch lengths in the vicinity of a topological rearrangement can be efficiently optimized via the Newton-Raphson method incorporating both the gradient and Hessian information for one branch at a time.
However, such optimization strategy is only efficient for optimization over a limited number of parameters,
because the computational complexity for evaluating the Hessian matrix increases quadratically with the number of parameters.

In the Bayesian analyses, our linear-time gradient algorithm allows efficient sampling of all branch-specific evolutionary rates from their posterior density using HMC.
The vanilla HMC sampler gains a 2.2- to 20.9-fold increase in learning the branch-specific rates with the minimum ESS per unit time criterion.
The preconditioning improves the efficiency of HMC
with a 16.4- to 33.9-fold increase.
The computational cost for evaluating the diagonal entries of the Hessian matrix is almost the same as the gradient (see Equation~\ref{Eq:Hessian}).
In fact, the first term is nearly identical to the gradient in Equation~\ref{Eq:GradientWithGamma} except for replacing the infinitesimal matrix $\mathbf{Q}_i$ and the discrete rate $\sr{l}$ by their quadratic forms.
The second term in Equation~\ref{Eq:Hessian} reuses the gradient evaluated at the current position from the cached values for updating the momentum (see Equation~\ref{eq:leapfrog}).
Moreover, we update the adaptive preconditioning mass matrix every $10$ iterations of the HMC sampler.
This limits the additional computational cost in evaluating the diagonal of the Hessian matrix.


We observe an inverse correlation between the variability of the scales among the branch-specific evolutionary rates and the spread of ESS per second for the `vanilla' HMC sampler as shown in Figure~\ref{fig:ESS}.
Specifically, using the standard deviation of the marginal posterior distribution as a qualitative measure for the scale, the West Nile virus, Lassa virus and Dengue virus examples return a variance across the standard deviations as $0.014$, $0.006$ and $0.036$ and the ratio between the maximum and the minimum of the standard deviations being $2.2$, $1.7$ and $17.8$ respectively.
The branch-specific evolutionary rates of the Dengue virus example exhibit the highest variability among the three datasets and the `vanilla' HMC sampler performs the worst for this dataset.
As discussed in Section~\ref{sec:preconditioning}, not accounting for high variability among the scales of the parameters reduces the efficiency of the `vanilla' HMC sampler.
Preconditioning improves the inadequate performance of the `vanilla' HMC sampler via the adaptive mass matrix informed by the diagonal elements of the Hessian.
The mass matrix incorporates the variation in scales among the branch-specific evolutionary rates with a negligible cost of additional computation.

\section{Acknowledgments}

We thank Jeffrey Thorne for thoughtful comments.
The research leading to these results has received funding from the European Research Council under the European Union's Horizon 2020 research and innovation programme (grant agreement no.~725422 - ReservoirDOCS).
The Artic Network receives funding from the Wellcome Trust through project 206298/Z/17/Z.
MAS and XJ are partially supported by NSF grant DMS 1264153 and NIH grants R01 AI107034 and U19 AI135995.
GB acknowledges support from the Interne Fondsen KU Leuven / Internal Funds KU Leuven under grant agreement C14/18/094.
PL acknowledges support by the Research Foundation -- Flanders (`Fonds voor Wetenschappelijk Onderzoek -- Vlaanderen', G066215N, G0D5117N and G0B9317N).




\clearpage
\bibliographystyle{imsart-nameyear}

\end{document}